\documentclass[journal]{IEEEtran}

\usepackage{diagbox}
\usepackage{graphicx}
\usepackage{caption}
\usepackage{float} 
\usepackage{subcaption}

\usepackage{makecell}

\ifCLASSINFOpdf
  
\else
  
\fi

\usepackage{amsmath}
\usepackage{algorithmic}
\usepackage{array}
\usepackage{multirow}
\usepackage{booktabs}

\usepackage{cite}
\usepackage{hyperref}     
\hypersetup{
    colorlinks=true,       
    linkcolor=blue,        
    citecolor=blue         
}

\usepackage{colortbl}
\usepackage[table]{xcolor}
\usepackage{makecell}
\begin{document}

 \title{A CGAN-LSTM-Based Framework for Time-Varying Non-Stationary Channel Modeling}

\author{Keying~Guo,
        Ruisi~He,~\IEEEmembership{Senior Member,~IEEE,}
        Mi~Yang,~\IEEEmembership{Member,~IEEE,}
        Yuxin~Zhang,~\IEEEmembership{Graduate Student Member,~IEEE,}
        Bo~Ai,~\IEEEmembership{Fellow,~IEEE,}
        Haoxiang~Zhang,
        Jiahui~Han,
        and~Ruifeng~Chen

\thanks{Part of this paper was submitted to the 2025 IEEE 102nd Vehicular Technology Conference (IEEE VTC2025-Fall).}

\thanks{K. Guo, R. He, M. Yang, Y. Zhang, B. Ai are with the School of Electronics and Information Engineering, Beijing Jiaotong University, Beijing 100044, China, and also with the State Key Laboratory of Advanced Rail Autonomous Operation, and the Frontiers Science Center for Smart High-speed Railway System, Beijing Jiaotong University, Beijing 100044, China (email: ky.guo@bjtu.edu.cn, ruisi.he@bjtu.edu.cn, myang@bjtu.edu.cn, yx.zhang@bjtu.edu.cn, aibo@ieee.org).}

\thanks{H. Zhang and J. Han are with China Academy of Industrial Internet, Ministry of Industry and Information Technology, Beijing, China (e-mail: zhx61778294@126.com, hjh1760708@126.com).}

\thanks{R. Chen is with the Institute of Computing Technology, China Academy of Railway Sciences Corporation Limited, Beijing 100081, China (email:ruifeng$\_$chen@126.com).}}

\maketitle

\begin{abstract}
Time-varying non-stationary channels, with complex dynamic variations and temporal evolution characteristics, have significant challenges in channel modeling and communication system performance evaluation.  Most existing methods of time-varying channel modeling focus on predicting channel state at a given moment or simulating short-term channel fluctuations, which are unable to capture the long-term evolution of the channel. This paper emphasizes the generation of long-term dynamic channel to fully capture evolution of non-stationary channel properties. The generated channel not only reflects temporal dynamics but also ensures consistent stationarity. We propose a hybrid deep learning framework that combines conditional generative adversarial networks (CGAN) with long short-term memory (LSTM) networks. A stationarity-constrained approach is designed to ensure temporal correlation of the generated time-series channel. This method can generate channel with required temporal non-stationarity. The model is validated by comparing channel statistical features, and the results show that the generated channel is in good agreement with raw channel and provides good performance in terms of non-stationarity. 
\end{abstract}

\begin{IEEEkeywords}
Dynamic channel modeling, non-stationary, wireless channels, CGAN, LSTM.
\end{IEEEkeywords}

\IEEEpeerreviewmaketitle

\section{Introduction}

\IEEEPARstart{W}{ith} the development of 5G and the forthcoming 6G technologies, application scenarios of wireless communications have expanded from traditional voice communication to high-demand use cases such as high-definition video, the Internet of things, vehicle-to-vehicle (V2V), and smart manufacturing. These developments have raised performance requirements of communication systems, including higher transmission rates, lower latency, and higher reliability\cite{9390169}. Achieving these goals relies heavily on accurate channel modeling, which forms the foundation of wireless communication system design\cite{9713745}. Channel modeling reveals the complex relationship between physical environment and the channel\cite{8570034,10466759,tian2025analytical}.

However, with the increasing complexity of communication environments, traditional channel modeling methods often fail to address the challenges faced in modern wireless communications\cite{9727146,he2021wireless}. In particular, non-stationarity of channels in high mobility, large antenna arrays, and complex environments has become a critical issue\cite{renaudin2010non}. Non-stationary channels are characterized by time, space, and frequency-dependent changes, and their statistical properties cannot remain constant at any given moment\cite{9318511,10342805,he2015characterization}. Such non-stationarity introduces uncertainty into system design and optimization, and directly impacts signal transmission quality, system capacity, energy efficiency, and network stability\cite{9485046}. For example, in dynamic scenarios like high-speed rail, V2V communication, and unmanned aerial vehicle communications, non-stationarity of channels can cause traditional communication protocols and optimization methods to fail, resulting in suboptimal communication quality or network efficiency\cite{6808529,5165328,8937764,he2024wireless}. Therefore, accurately modeling and simulating non-stationary channels is critical for ensuring efficient operation of wireless communication systems.

Current channel modeling methods mainly rely on two approaches: geometrical modeling and stochastic modeling\cite{10466759}. Geometrical modeling describes channel characteristics based on physical environments, providing high accuracy but with large computational overhead, making it unsuitable for time-varying non-stationary channels\cite{8114238,molisch2003geometry}. In contrast, stochastic modeling uses probability distributions to characterize channel variations, which is computationally efficient but struggles to accurately simulate complex physical propagation paths, especially in highly dynamic environments\cite{9714331}. Additionally, traditional channel measurement methods face limitations such as high costs, experimental complexity, and an inability to cover all potential scenarios or frequency bands comprehensively\cite{1696471,9810138,zhao2018wireless}.

\subsection{Related Work}
Against this backdrop, artificial intelligence (AI) based approaches have emerged as promising alternatives for channel modeling, which can compensate for the limitations of traditional channel modeling methods\cite{he2024cost}. In particular, deep learning technologies have shown potential in learning channel characteristics and generating channel data. AI-based channel modeling can be used for channel parameter estimation\cite{9713743}. In \cite{8786170}, an artificial neural network (ANN) model was developed to predict channel excess attenuation in Q-band satellite communication, leveraging weather conditions and previous state data for fading estimation. Ref.\cite{9336252} studied a machine-learning-based method using support vector machine (SVM) and principal component analysis (PCA), which achieved real-time angle-of-arrival recognition in dynamic vehicular communication environments. 

AI method has a strong ability to recognize patterns in data and meets the requirements of cluster recognition. Ref.\cite{8013075} proposed a kernel-power-density (KPD)-based algorithm multipath component (MPC) clustering. In \cite{8555662}, a target recognition based clustering algorithm was developed for time-varying channels. The clusters in power angle spectrum extracted from measurement data are separated from the background. 
The high dimensionality and abstraction of the feature space of wireless channels motivates the use of generative adversarial network (GAN) to generate channels and expand channel datasets. Ref.\cite{8663987} proposed a novel GAN framework to address the problem of autonomous channel modeling without complex theoretical analysis or data processing. In \cite{9669188}, a DL-based channel modeling and generating approach namely ChannelGAN was proposed, designed on a small set of 3rd generation partnerships project (3GPP) link-level multiple-input multiple-output (MIMO) channel. Ref.\cite{10238401} achieved a channel modeling based on GANs, which can generate identical statistical distribution with measured channel.

In addition to this, existing studies have paid attention to dynamic characteristics of channel. recurrent neural networks (RNN) contribute to study of spatial and temporal correlation of wireless channels. The work in \cite{10466759} introduced a frequency domain predictive channel model that combines an autoencoder with a coupling convolution gated recurrent unit (Conv-GRU) cells. The proposed model aims to predict channel characteristics in unknown frequency bands. Ref.\cite{10417075} leveraged convolutional time-series generative adversarial network (TimeGAN) to capture the intrinsic features of the original datasets and then generate synthetic samples. A GAN and LSTM based channel prediction framework was proposed in \cite{9860457} to model indoor wireless channels, enriching channel data and enabling sequence prediction. Ref.\cite{ 10461110} is a further study of \cite{9860457}, presenting a novel 6G space-time joint predictive channel model to predict channels in space-time domains. Ref.\cite{9676455} investigated a convolutional neural network (CNN) and convolutional long short-term memory (CLSTM) based channel prediction model by considering temporal and spatial correlations of massive MIMO channels in high-speed railway. Although these studies have made significant advances in spatial and temporal correlation and channel prediction, they often face limitations. Many models focus on short-term channel fluctuations and fail to effectively capture non-stationary characteristics of long-term channel evolution. On the other hand, existing modeling approaches often assume some degree of stationarity at a given time, which is inadequate for handling strong non-stationary channel variations.  Moreover, while AI-based methods have shown promise in generating channel , the lack of comprehensive modeling of time-domain non-stationarity means that the generated channel data often lacks consistency and stability required for high-precision communication systems over long time scales. 
\subsection{Contributions}
To the best of the authors’ knowledge, there is currently a lack of research on acquiring long-term continuous-time dynamic channel data to fully capture evolution of channel properties over time and ensuring stationarity consistent with raw channel. To fill the gaps, based on the preliminary work presented in \cite{gky}, this paper proposes a hybrid deep learning architecture combining CGAN and LSTM networks to generate dynamic channel data over a period of time. The proposed method captures temporal evolution of channels whereas ensuring that the generated data aligns with the stationary properties of simulation data. By introducing stationarity constraints in generator, this approach achieves high-precision dynamic channel modeling, producing data with statistical properties close to raw channels. The main contributions of this work are as follows:
\begin{itemize}
	\item A hybrid CGAN-LSTM architecture is proposed for modeling dynamic non-stationary channels, aiming to generate time-domain non-stationary dynamic channel data that aligns with raw channel characteristics. The architecture combines CGAN with LSTM networks, enabling it to model both non-stationary channel characteristics and the long-term temporal correlation. 
	\item To enhance physical consistency of the generated channels, an innovative stationarity constraint mechanism is introduced. This mechanism ensures temporal correlation of the generated channels, preventing non-physical fluctuations in channel evolution and maintaining high consistency and realism throughout the dynamic process.
	\item The effectiveness of the proposed method is validated in V2V scenarios, offering a high-precision generative solution for channel modeling in complex scenarios. 

\end{itemize}

This paper is organized as follows. Section II introduces the proposed CGAN-LSTM channel modeling framework. Section III presents experimental implementation. Next, experimental results and analysis are demonstrated in Section IV. Finally, Section V concludes the paper.

\section{Channel Modeling Framework}

In this section, we propose a novel CGAN-LSTM-based channel modeling framework aiming at generating channel data with accurate non-stationary characteristics. The framework contains the architecture design, which includes generator and discriminator network design and embedding LSTM into the generator. In addition, an auxiliary supervision module is proposed to design correlation loss function to ensure that the generated channel data have similar non-stationarity.

\begin{figure*}[!t]
\centering
\includegraphics[width=7in]{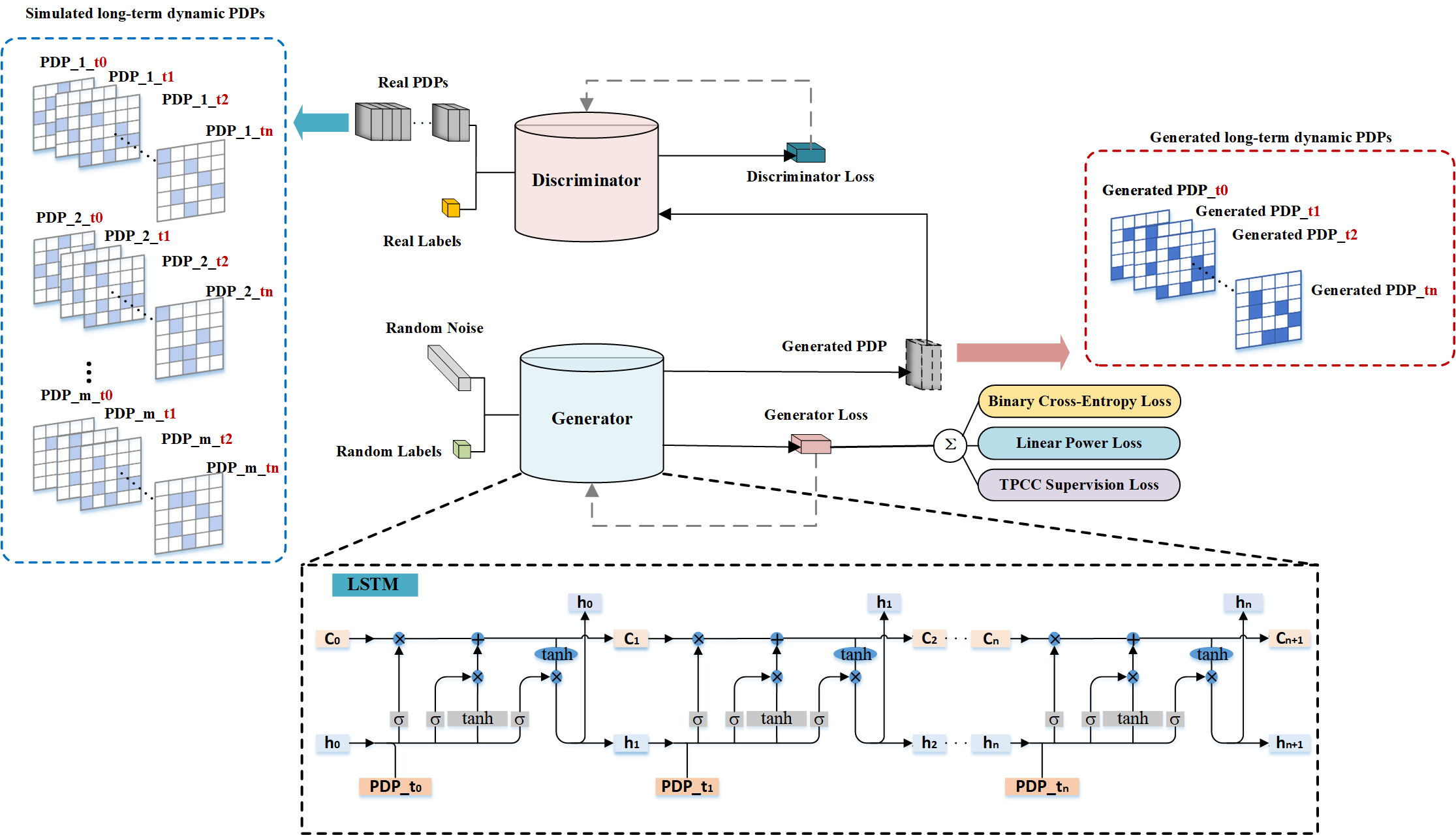}
\caption{Network structure of the CGAN-LSTM model with temporal stationarity constraints.}
\label{network}
\end{figure*}
\subsection{Network Architecture}

In this paper, power delay profile (PDP) is used as the target for modeling channel data. Both training and generated channels are represented in the form of PDPs to capture the dynamic characteristics of the channel. The primary parts of the network lie in accurately capturing and replicating inherent non-stationary dynamics of raw channels, such as snapshot-to-snapshot correlations and time-varying statistical distributions. To address  these issues, the framework incorporates three key components: generator, discriminator, and auxiliary supervision module. The generator leverages CGAN principles and integrates LSTM layers to capture and maintain temporal correlations in the generated PDPs of channel. By learning from labeled training data, the generator not only synthesizes realistic PDP samples but also models temporal evolution of channel characteristics. The discriminator, acting as a classification network, evaluates authenticity of PDP samples and provides adversarial feedback to the generator. This process ensures that the generated PDPs align closely with the statistical characteristics of training data. The auxiliary supervision module is embedded within the generator to  maintain snapshot-to-snapshot temporal correlations. This module imposes snapshot similarity constraint during training process, directly guiding generator to produce PDPs that exhibit the required temporal stability.

\subsubsection{CGAN Model}
CGAN is an extension of the traditional GAN framework, where both generator and discriminator are conditioned on additional information, such as labels or auxiliary data\cite{ding2021ccgan}. This conditioning enables CGAN to generate data under specific conditions. Core components of CGAN are the generator and the discriminator. The generator is trained to produce PDPs that closely matches the distribution of raw channel PDPs, whereas the discriminator is trained to distinguish between raw and generated PDPs. Both networks operate under the guidance of an additional input vector, typically representing a label or a specific condition.

The generator's goal is to minimize the loss function by improving its ability to generate PDPs that the discriminator misclassifies as real. Simultaneously, the discriminator's objective is to maximize the same loss function by correctly classifying real and generated PDPs. The adversarial training process alternately updates generator and discriminator parameters to achieve equilibrium, where generator produces highly realistic PDPs, and discriminator can no longer distinguish between real and generated PDPs.

\subsubsection{LSTM Model}
LSTM is a type of RNN designed to learn temporal dependencies in sequential data. Key component of an LSTM unit is cell state, which carries information across time steps, and three gates: forget gate, input gate, and output gate. These gates decide what information is dropped, what is updated, and what is output at each time step. Moreover, using gate mechanism to control flow of information and mitigate issues related to vanishing and exploding gradients makes it particularly suitable for modeling non-stationary channel data with long-term dependencies\cite{6789445}. The operations of an LSTM cell are defined as follows:

Forget gate determines which information from the previous cell state $\tilde{C}_{t-1}$ should be discarded. It is defined as: 
\begin{equation}\begin{aligned}
f_t=\sigma\left(W_f\cdot[h_{t-1},x_t]+b_f\right),
\end{aligned}\end{equation}
where $\sigma$ denotes the sigmoid function, $W_f$ and $b_f$ are the weights and biases of forget gates, $h_{t-1}$ is hidden state at the previous moment, and $x_t$ is input at the current moment.

Input gate determines which new information should be added to the cell state. It is defined as: 
\begin{equation}\begin{aligned}
 i_t=\sigma\left(W_i\cdot[h_{t-1},x_t]+b_i\right) ,
\end{aligned}\end{equation}
where $W_i$ and $b_i$ are the weights and biases of input gates.

Simultaneously, a candidate cell state is created to represent the new potential information to be added:
\begin{equation}\begin{aligned}
 \tilde{C}_t\:=\tanh\left(W_C\cdot\left[h_{t-1},x_t\right.\right]+b_C\:) ,
 \end{aligned}\end{equation} 
and the actual cell state $C_t$ is then updated by combining the results of the forget gate and the input gate: 
 \begin{equation}\begin{aligned}
C_t=f_t\cdot C_{t-1}+i_t\cdot\tilde{C_t},
 \end{aligned}\end{equation} 
 where $C_{t-1}$ is cell state at the previous moment, and $W_C$ and $b_C$ are weights and biases of the cell state.

 The output gate determines what part of the cell state should contribute to the hidden state $h_t$ at the current time step. It is defined as: 
 \begin{equation}\begin{aligned}
o_t=\sigma\left(W_o\cdot[h_{t-1},x_t]+b_o\right),
 \end{aligned}\end{equation}
 and the current hidden state is computed as:
  \begin{equation}\begin{aligned}
h_t=o_t\cdot\tanh\left(C_t\right),
 \end{aligned}\end{equation}
where $W_o$ and $b_o$ are the weights and biases of the output gates, and $h_t$ is hidden state at the current moment in time. These mechanisms allow the network to effectively capture evolving patterns in dynamic channels. 

\subsubsection{Combining CGAN and LSTM}
In Fig. \ref{network}, network structure of the CGAN-LSTM model with temporal stationarity constraints is illustrated. The model employs a combination of CGAN and LSTM networks to generate PDPs that reflect dynamic characteristics and temporal correlations of wireless channels. Multiple groups of long-term dynamic channel PDPs are fed into the network, in which $PDP\_m\_tn$ represents the PDP slice of the $m$-th dynamic channel at time $tn$. LSTM network is embedded in the generator, where each LSTM state corresponds to a time slice in the dynamic channel. By incorporating LSTM, the model learns the temporal dynamic characteristics of channel, specifically capturing correlations between consecutive time slices or snapshots. These dependencies are closely related to non-stationarity of channel: stronger correlations between snapshots indicate higher channel stationarity, whereas weaker correlations reflect higher non-stationarity. Thus, stationarity of channel can be associated with the temporal correlation between snapshots, which serves as a quantifiable metric for characterizing channel stationarity.

To evaluate the modeling and generation performance under varying degrees of channel stationarity, two categories of channels are designed: strongly non-stationary channels and weakly non-stationary channels. These categories enable comprehensive analysis and validations of the proposed method’s ability to capture and reproduce dynamic channel characteristics across different levels of non-stationarity. The CGAN framework ensures that the generated samples are conditioned on appropriate labels, representing distinct channels with strong and weak non-stationarity. This conditional approach enables the model to generate realistic PDPs that not only replicate the temporal characteristics of the training data but also maintain consistency with the associated non-stationarity levels. By integrating these components, the framework achieves a balance between accurate replication of the underlying channel dynamics and preserving the non-stationarity characteristics, ensuring the generated PDPs closely resemble the behavior of raw wireless channels.

\subsection{Loss Function Design}
To further ensure accuracy of the generated channel loss function is designed to incorporate three key components: binary cross-entropy loss, line loss function for linear power, and temporal PDP correlation coefficient (TPCC) loss. The binary cross-entropy loss is employed to guide the generator in producing outputs that closely resemble raw channels, enhancing authenticity of the generated PDP. The linear power loss ensures that the generated PDP maintain correct power distribution, thereby preserving physical characteristics of channel. The WSS loss is introduced to explicitly address non-stationary nature of wireless channel. Since non-stationarity of channel is a critical characteristic, particularly in dynamic environments, this loss function penalizes discrepancies in temporal correlation and stationarity of the generated PDPs relative to the training data. 
\subsubsection{Binary Cross-Entropy Loss}
The adversarial training process of CGAN relies on a binary cross-entropy loss function. For the discriminator, the goal is to minimize the loss associated with misclassifying raw PDP samples \( P \) and generated PDP samples \(\hat{P}\). For the generator, the aim is to maximize likelihood of discriminator being fooled by \(\hat{P}\). The binary cross-entropy loss function is defined as follows:

\begin{equation}
\begin{aligned}
L_{D} = & -\dfrac{1}{2} \Big( E_{P \sim p_{\text{real}}(P)}[\log D(P \vert y)] \\
& + E_{z \sim p_{z}(z)}[\log(1 - D(G(z \vert y), y))] \Big),
\end{aligned}
\end{equation}

\begin{equation}
\begin{aligned}
L_{G} = -E_{z \sim p_{z}(z)}[\log(D(G(z \vert y), y))],
\end{aligned}
\end{equation}
where \( L_{D} \) and \( L_{G} \) represent binary cross-entropy loss functions of discriminator and generator, respectively. \( P \) denotes real data drawn from raw data distribution \( p_{\text{real}}(P) \), \( y \) is the conditional label or auxiliary information, \( z \) is latent variable (random noise), \( G(z \vert y) \) represents the generated data given conditional label \( y \), and \( \hat{P} = G(z \vert y) \) denotes the generated sample. \( D(P \vert y) \) is the probability that discriminator assigns to real sample \( P \), conditioned on \( y \), and \( D(\hat{P} \vert y) \) is the probability that discriminator assigns to the generated sample \(\hat{P}\), conditioned on \( y \).

\subsubsection{Line Loss Function for Linear Power}
To ensure that the generated PDPs are accurate in terms of linear power distribution, we introduce a line loss function that operates in linear power domain. The loss function is calculated as follows:

\begin{equation}
% L_{line}=\dfrac{1}{N}\sum_{i=1}^{N}(\frac{P_{i} -\min ({P})   }{\max ({P}_{i} ) -\min ({P}_{i} ) } -\frac{\hat{P}_{i} -\min (\hat{P}_{i} )   }{\max (\hat{P}_{i} ) -\min (\hat{P}_{i} ) })  ^2
L_{linear}= \sum_{n=1}^{N}\sum_{m=1}^{M}(\frac{P(t_{m} ,\tau_{n}) -\hat{P}(t_{m} ,\tau_{n})   }{\max ({P}) -\min ({P} ) } )  ^2,
% L_{line}= \int \int (\frac{P(t_{i} ,t_{j}) -\hat{P}(t_{i} ,t_{j})   }{\max ({P}) -\min ({P} ) } )  ^2dt\ d\tau\
\end{equation}
where $t$  represents time, $\tau$  represents delay, $M$ and $N$ are numbers of time and delay points, $P(t_{m} ,\tau_{n})$ and $ \hat{P}(t_{m} ,\tau_{n})$ represent the rawl and generated linear power at  time $t_{m}$ and delay $\tau_{n}$. This line loss enforces that the generated data closely matches the linear power distribution of training data.
\subsubsection{TPCC Supervision Loss}
The introduction of TPCC supervision loss aims to ensure consistency of temporal correlation between consecutive snapshots in the generated channels. In this study, TPCC is used as a metric for channel stationarity, which helps to distinguish and quantitatively measure non-stationarity of strong and weak non-stationary channels. According to the definition provided in \cite{gehring2001empirical}, the TPCC between PDPs at time moments $t_{i}$ and $t_{j}$ can be calculated as:
\begin{equation}
c(t_{i} ,t_{j})= \dfrac{\int P(t_{i} ,\tau)\cdot P(t_{j} ,\tau) d\tau  }{\max \{ \int ^{2} P(t_{i} ,\tau)d\tau, \int ^{2} P(t_{j} ,\tau)d\tau\} },
\end{equation}
where $P(t_{i} ,\tau)$ and $P(t_{j} ,\tau)$ represent powers at time $t_{i}$ and $t_{j}$.
The TPCC is normalized to lie between 0 and 1. A threshold is then set, and the TPCC values exceeding this threshold are counted, providing a quantitative measure of channel temporal non-stationarity. This allows us to further compute the generalized stationary interval by using simple relationships between time, distance, and velocity. The stationary time represents quantification of channel stationarity in time domain.

The TPCC loss function is defined to penalize discrepancies in temporal correlations between consecutive snapshots, encouraging the generator to produce more realistic and temporally consistent channel realizations. The TPCC loss can be formulated as:
\begin{equation}
L_{\mathrm{TPCC}}=\frac{1}{M}\sum_{i=1}^M\sum_{j=i+1}^M|\mathrm{TPCC}(t_i,t_j)-\hat{\mathrm{TPCC}}(t_i,t_j)|,
\end{equation}
where $\mathrm{TPCC}(t_i,t_j)$ and $\hat{\mathrm{TPCC}}(t_i,t_j)$ represent TPCC values between $t_{i}$ and $t_{j}$ calculated between the raw and generated PDPs. 
\subsubsection{Final Generator Loss}
The final loss function for the generator combines binary cross-entropy loss, line loss for linear power, and TPCC supervision loss. The total generator loss is computed as:
\begin{equation}
L_{\mathrm{gen}}=\lambda_1L_{\mathrm{G}}+\lambda_2L_{\mathrm{linear}}+\lambda_3L_{\mathrm{TPCC}},
\end{equation}
where $\lambda_1$, $\lambda_2$ and $\lambda_3$ are the hyperparameters controlling the weights of $L_{\mathrm{G}}$, $L_\mathrm{linear}$ and $L_\mathrm{TPCC}$.
By incorporating these three losses into generator objective function, the model is able to learn fundamental features of channel, and effectively captures, non-stationary behaviors in the training data. The comprehensive loss function thus enhances the generator to produce realistic and non-stationary channel realizations.

\begin{figure}[!t]
\centering
\begin{subfigure}[t]{0.6\linewidth}
    \centering
    \includegraphics[width=\linewidth]{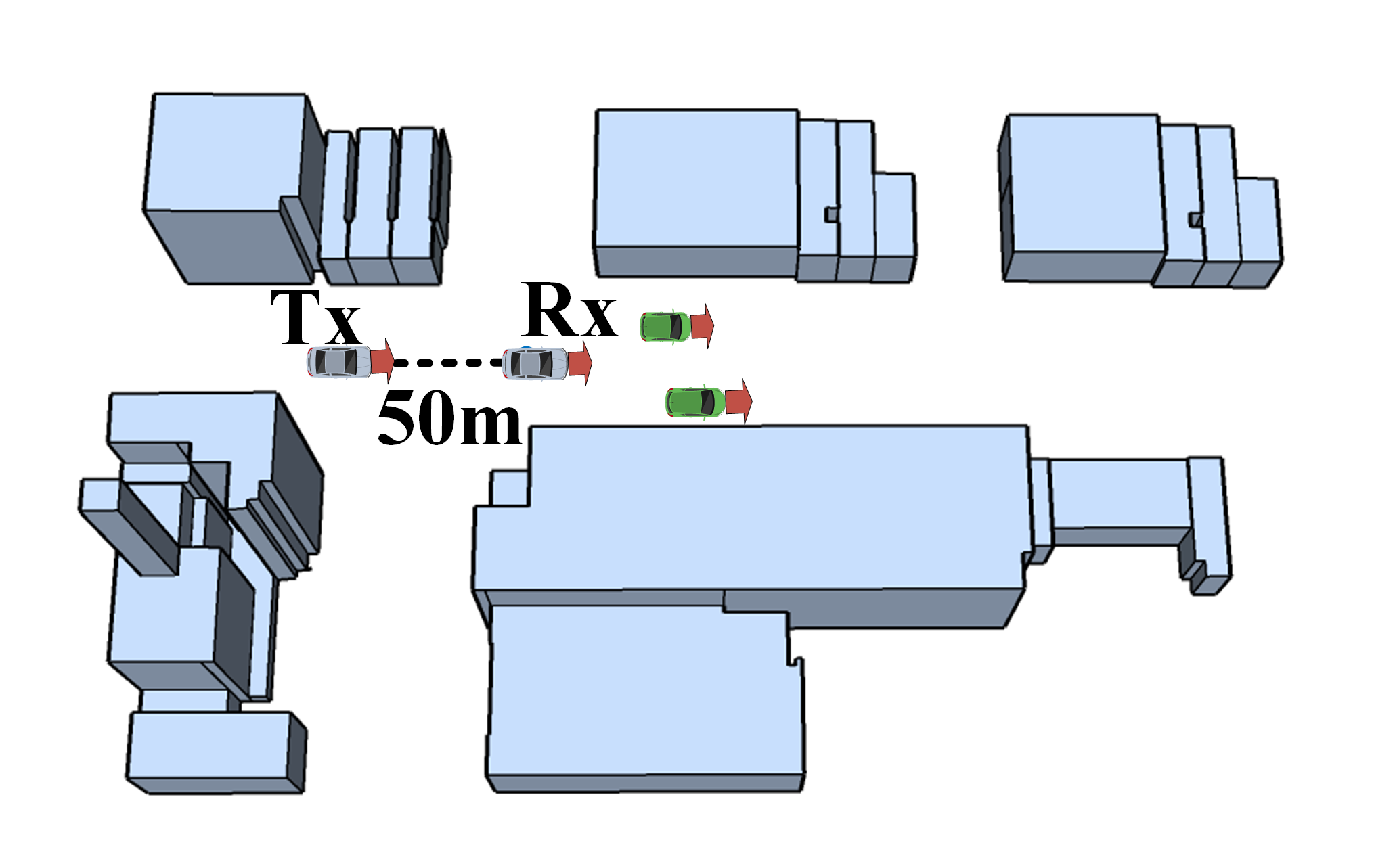}%{e2.PNG}
    \caption{}
\end{subfigure}
\vspace{0.1cm}
\begin{subfigure}[t]{0.9\linewidth}
    \centering
    \includegraphics[width=\linewidth]{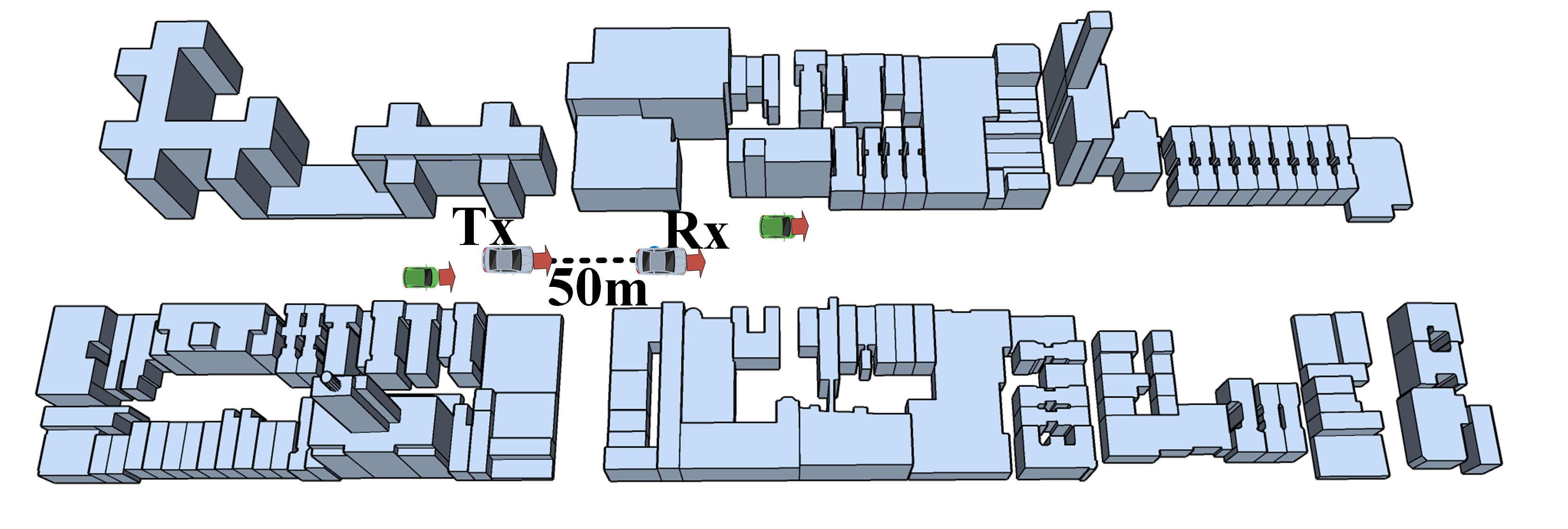}%{d2.png}
    \caption{}
\end{subfigure}

\caption{Examples of RT simulation scenario layouts: (a) one of the sparse scenario layouts, (b) one of the dense scenario layouts.}
\label{Simu}
\end{figure}

\section{Model Implementation}
In this section, implementation of the proposed model is presented. This includes description of dataset generated by  ray tracing (RT) simulation. The dataset is used for training and evaluating performance of the proposed model. In addition, network architecture and parameter settings for experiments are described.
\subsection{Channel Dataset}

The training dataset is generated using the RT simulation platform based on \cite{8663551}. To investigate the channels with varying levels of non-stationarity, two categories of channels are designed to represent different levels of non-stationarity: weakly non-stationary channels with sparse scatterers and low speed, and strongly non-stationary channels with dense scatterers and high speed. This classification emphasizes the impact of scatterer density and mobility on dynamic properties of channels. To model the dynamic channel characteristics, V2V channel simulations are conducted.

\begin{itemize}
	\item Category 1: Weakly non-stationary channels. This category features sparse scatterers distributed on both sides of street. The Tx and Rx vehicles move uniformly in the same direction at a low speed of 0.5 m/s, maintaining a fixed distance of 50 meters and a height of 1.5 meters, positioned in a designated lane. Two dynamic scatterer vehicles are randomly distributed in other lanes, moving at the same speed and direction as Tx and Rx. Their initial positions are randomly assigned along the road direction within a range of 5 to 25 meters from Tx and Rx. To ensure integrity of line-of-sight (LOS) path, dynamic scatterers are strictly confined outside the LOS region, introducing only minor variations in multipath components, resulting in weakly non-stationary channel characteristics.

        \item Category 2: Strongly non-stationary channels. This category features dense scatterers distributed on both sides of street. Tx and Rx move uniformly in the same direction at a higher speed of 1.5 m/s, maintaining a fixed distance of 50 meters and a height of 1.5 meters. The placement and rules for the two dynamic scatterer vehicles are identical to those in Category 1. However, the increased scatterer density and higher mobility result in more significant multipath effects, leading to strongly non-stationary channel characteristics.
\end{itemize}

Both Tx and Rx are equipped with omni-directional antennas, operating at a center frequency of 6 GHz with a bandwidth of 150 MHz. Each category includes three distinct scenario layouts, and 300 independent dynamic simulations are performed for each scenario layout. Fig.\ref{Simu} shows examples of simulation scenario layouts. These configurations introduce diversity to the dataset while ensuring statistical consistency within each category. A total of 1800 independent dynamic channels are generated, with each channel containing 300 consecutive snapshots to capture temporal evolution. The dataset is thus structured as [1800, 300, 300], where 1800 denotes total number of dynamic channels, the first 300 represents number of snapshots per channel, and the second 300 corresponds to frequency points per snapshot.

Based on the above simulations, the obtained dataset exhibits distinct non-stationary characteristics. As suggested in \cite{8851421,9485046}, the wide-sense stationary (WSS) region is used to evaluate channel stationarity. Channels with larger WSS regions demonstrate higher stationarity, while smaller WSS regions correspond to highly non-stationary channels with rapid temporal variations. In dense scatterer scenarios, the abundance of reflectors and scatterers leads to richer multipath effects and larger RMS delay spreads (RMSDS), indicating greater temporal dispersion. Conversely, sparse scatterer scenarios yield fewer multipaths, resulting in smaller RMSDS and relatively stable channel characteristics. To ensure consistency within each category, the simulation results are filtered to maintain similar channel characteristic distributions across all scenarios in the same category.

\subsection{Data Pre-Processing}
In order to improve quality of training data and adapt input to network requirements, we implement a data pre-processing step that includes masking and normalization. This step ensures that outlier values and invalid regions do not negatively impact model training, as suggested in \cite{garcia2016big}.
\subsubsection{Masking Process}
 Masking is specifically used to filter out invalid or physically irrelevant data in training set\cite{Suvorov_2022_WACV}. For the received power below the predetermined threshold, these data treated as noise or physically unusable information are identified and replaced with threshold value to prevent extreme values from influencing training process. By applying a mask, this data is set to a fixed value and effectively excluded from learning process. In this way, the network is able to focus on modeling data that is closely related to raw channel characteristics, reducing noise interference and improving accuracy of the generated output dynamics. A binary mask is created to differentiate between valid and masked regions:
\begin{equation}
\mathrm{Mask}(i,j)=\begin{cases}
1, & \mathrm{if~}P(i,j)\geq T_{\mathrm{threshold}}, \\
0, & \text{otherwise.}
\end{cases}
\end{equation}
where $T_{\mathrm{threshold}}$ is a set threshold, $P(i,j)$ is raw power of input, and ${Mask}(i,j)$ denotes mask matrix corresponding to the input power. The mask corresponding to data less than threshold is set to 0, and mask for data above threshold is set to 1. The data are then updated as follows:
\begin{equation}
P(i,j)=\begin{cases}
P(i,j),& \text{if~}Mask(i,j)=1,\\
T_{\text{threshold}},& \text{if~}Mask(i,j)=0.
\end{cases}
\end{equation}
This masking operation ensures that values below threshold are preserved without distortion and marked as special regions for potential differentiation during training.
\subsubsection{Normalization Process}
Data normalization standardizes feature distribution during preprocessing. Given that the training data may exhibit significant numerical disparities, for example, differences in magnitudes of power and path delay can lead to gradient instability or vanishing during training. By normalizing data to a uniform range, training convergence is accelerated, and activation functions operate in a more optimal regime, enhancing network nonlinear modeling capabilities and ensuring consistency in the distribution of generated PDP. The data is normalized to $[-1,1]$ to ensure consistent scaling and better convergence during training:
\begin{equation}
P_{{\text{normalized}}}=2\cdot\frac{P-P_{{\mathrm{min}}}}{P_{{\mathrm{max}}}-P_{{\mathrm{min}}}}-1,
\end{equation}
where $P_{{\mathrm{min}}}$ and $P_{{\mathrm{max}}}$ are the minimum and maximum values of $P$.
\subsubsection{Application of the Mask}
Normalized data can be more accurately mapped to the target generation range, and combination of normalization and masking enables the network to efficiently learn and reconstruct  critical dynamic channel characteristics.The masked regions are assigned a fixed normalized value of to differentiate them from other data:
\begin{equation}
P_{\mathrm{mask}}(i,j)=\begin{cases}
P_{\text{normalized}}(i,j),& \text{if~}Mask(i,j)=1,\\
-1,& \text{if~}Mask(i,j)=0.
\end{cases}
\end{equation}
This approach allows the network to learn valid data regions whereas still accounting for masked areas as special cases.

\begin{table}[!t]
% \normalsize
\small
    \centering
    \caption{Parameter Setting in Model Design and Training.}
    \label{Parameter setting}
    \begin{tabular}{l|c}
    \hline
    \hline
       Parameter&  Value \\
        \hline
       Number of LSTM &2\\
       $\alpha$ in Leaky ReLU & 0.2  \\
        Momentum of BatchNormalization & 0.8 \\
        Learning Rate of Adam & 0.0004\\
        Betas of Adam & 0.8 \\
        Dropout & 0.4 \\
        Number of Classes & 2 \\
        \hline
        \hline
    \end{tabular}
\end{table}

\subsection{Implementation}
In this section, we implement CGAN-LSTM for time-varying channel modeling. The generator is used to synthesize channel PDP, whereas discriminator evaluates veracity of these generated samples.
The generator takes as input a latent vector of size latent dimension along with a label as a condition for data generation. Its architecture starts with three fully connected layers of sizes 2048, 1000, and [300,300], which transform input into a format that can be reshaped into a two-dimensional matrix representing snapshots and delay boxes. In order to capture temporal correlation between snapshots, two stacked LSTM layers are applied separately. The output is further processed through a fully connected layer with tanh activation function to generate channel PDP that matches the shape of training samples. A label embedding mechanism is incorporated into the generator, where labels are embedded and combined with latent vectors through elementwise multiplication.

The discriminator takes the generated and simulated dynamic channels and their corresponding labels as inputs. Both inputs are flattened and combined through element-wise multiplication after embedding the labels into a space matching the dimensions of input data. The discriminator uses three fully connected layers of dimensions 2048, 1024, and 512, with LeakyReLU activations and Dropout layers to prevent overfitting. The final output is obtained through a sigmoid activation, representing the probability that input is real or generated. Parameter settings are presented in Table \ref{Parameter setting}.

\renewcommand{\arraystretch}{2} 
\setlength{\arrayrulewidth}{0.8pt} 
\begin{table*}[!htp]

\scriptsize 
    \centering
    \caption{Channel Statistics Parameters.}
    \label{channel statistics parameter}
    \begin{tabular}{c|c|c|c|c|c|c|c|c}
    \hline
    \hline
       \multirow{2}*{\makecell{ \textbf{Mean Value} }}& \multicolumn{4}{c|}{\textbf{Weakly Non-stationary Scenarios}}  & \multicolumn{4}{c}{\textbf{Strongly Non-stationary Scenarios}} \\
       \cline{2-9}
		~ & Simulation & CGAN-LSTM & \makecell{Undesigned \\CGAN-LSTM}&CGAN-GRU& Simulation & CGAN-LSTM & \makecell{Undesigned \\CGAN-LSTM} &CGAN-GRU  \\
        \hline
     \textbf{WSS interval (s)} & 8.88 &  8.51&6.16&8.54   &2.94&2.62& 2.10&2.65\\
       \textbf{RMSDS (ns)}& 26.74&26.38&25.20&24.81   &40.48&40.45&37.79&38.65 \\
       \textbf{Multipath Count}&62.72&65.94&66.78&59.87    &149.29&149.22&149.05&147.10\\
       \textbf{Shadow Fading (dB)}& -0.017& -0.070&-0.400&-0.214   &-0.012&-0.069& -0.426&-0.283\\
       \textbf{Path Loss (dB)}&83.98&83.94& 83.03&83.41   &83.99& 84.07&83.78& 83.64\\
        \hline
        \hline
    \end{tabular}
\end{table*}

\section{Results And Analysis}
\begin{figure}[!t]
\centering
\begin{subfigure}[t]{0.49\linewidth}
    \centering
    \includegraphics[width=\linewidth]{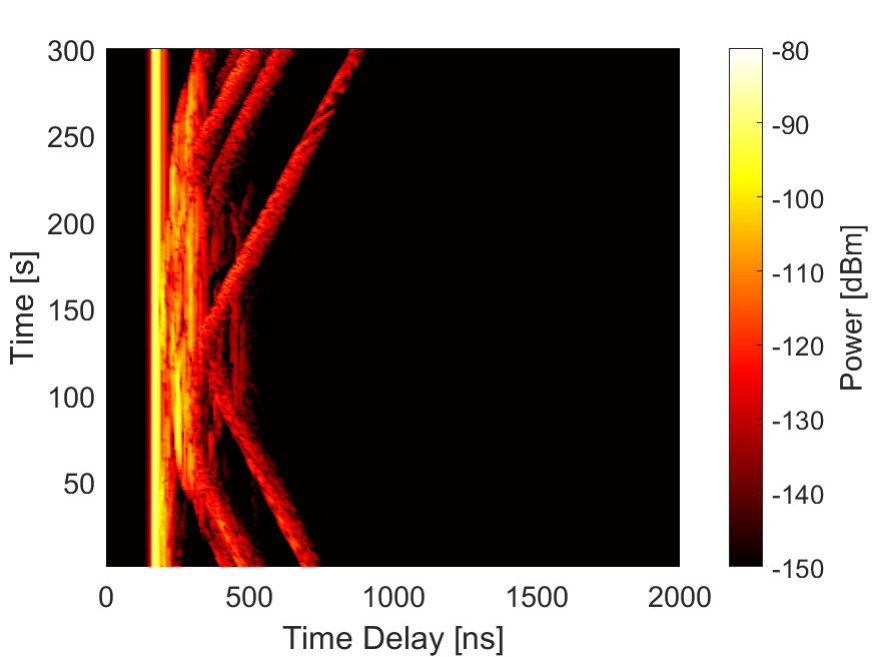}
    \caption{}
\end{subfigure}
\hfill
%\hspace{-5mm}
\begin{subfigure}[t]{0.49\linewidth}
    \centering
    \includegraphics[width=\linewidth]{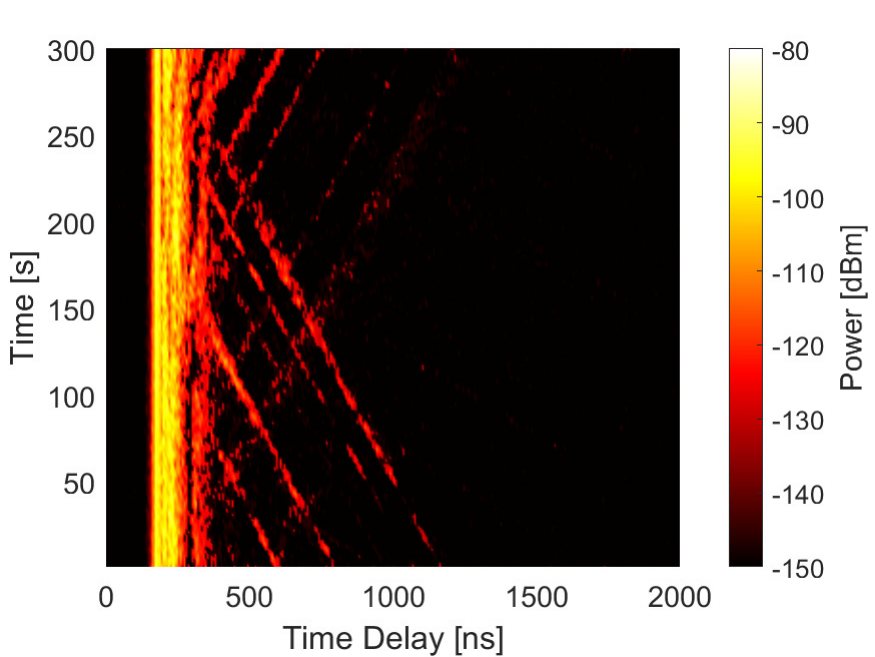}
    \caption{}
\end{subfigure}

\vspace{0.1cm}

\begin{subfigure}[t]{0.49\linewidth}
    \centering
    \includegraphics[width=\linewidth]{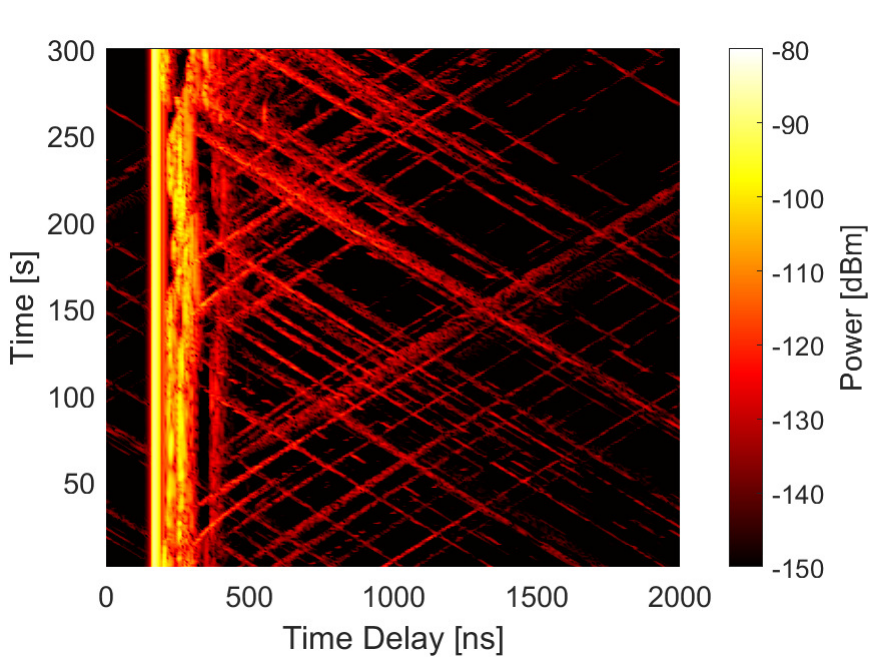}
    \caption{}
\end{subfigure}
\hfill
\begin{subfigure}[t]{0.49\linewidth}
    \centering
    \includegraphics[width=\linewidth]{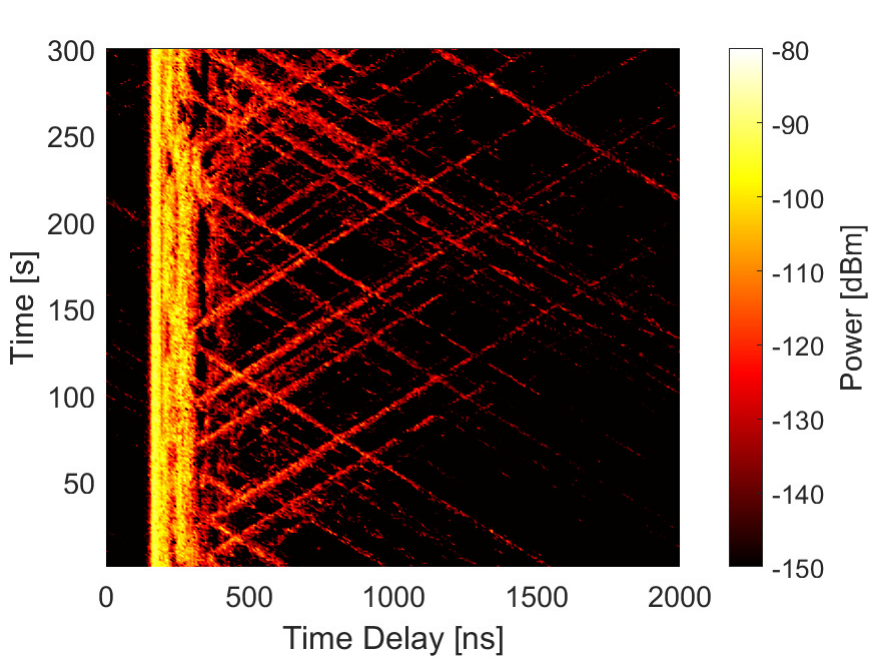}
    \caption{}
\end{subfigure}
\caption{Examples of RT simulated PDPs and the generated PDPs based on the proposed CGAN-LSTM in sparse and dense scenarios: (a) RT simulated the PDP in sparse scenarios, (b) CGAN-LSTM generated PDP in sparse scenarios, (c) RT simulated PDP in dense scenarios, and (d) CGAN-LSTM generated PDP in dense scenarios.}
\label{PDP}
\end{figure}
\begin{figure}[!t]
\centering
\begin{subfigure}[t]{0.9\linewidth}
    \centering
    \includegraphics[width=\linewidth]{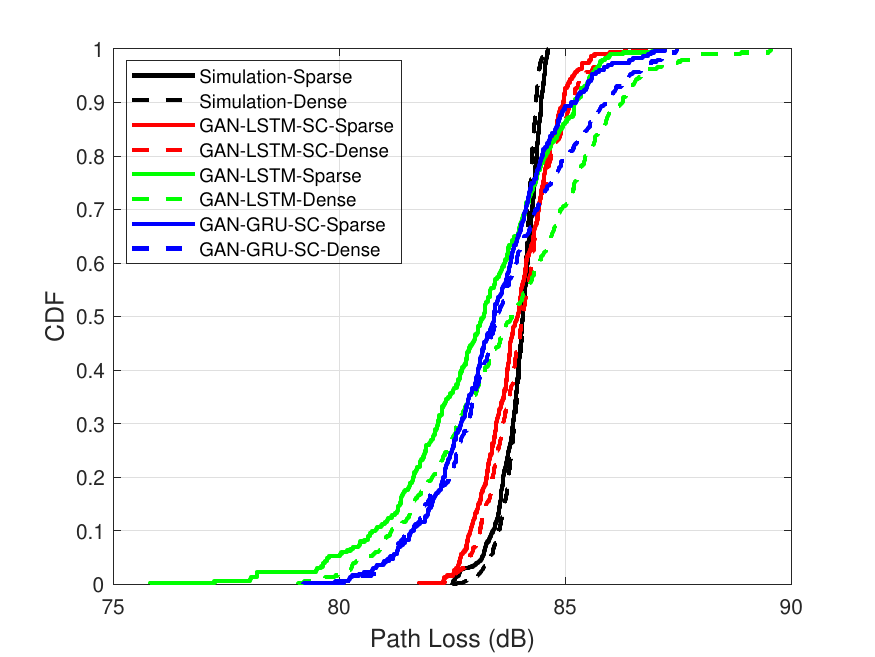}
    \caption{}
    \label{PL}
\end{subfigure}
\hspace{-5mm}
\begin{subfigure}[t]{0.9\linewidth}
    \centering
    \includegraphics[width=\linewidth]{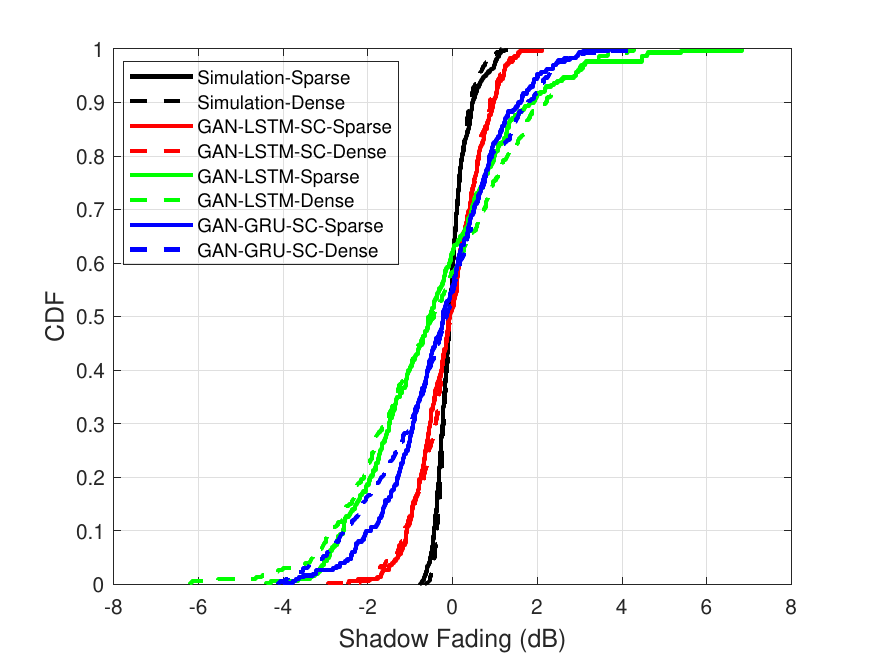}
    \caption{}
    \label{SF}
\end{subfigure}
\caption{Comparison of the distribution of fundamental statistical characteristics in sparse and dense scenarios:(a) shadow fading, and (b) path loss.}
\label{ALL2}
\end{figure}

\begin{figure}[!t]
\centering
\begin{subfigure}[t]{0.5\linewidth}
    \centering
    \includegraphics[width=\linewidth]{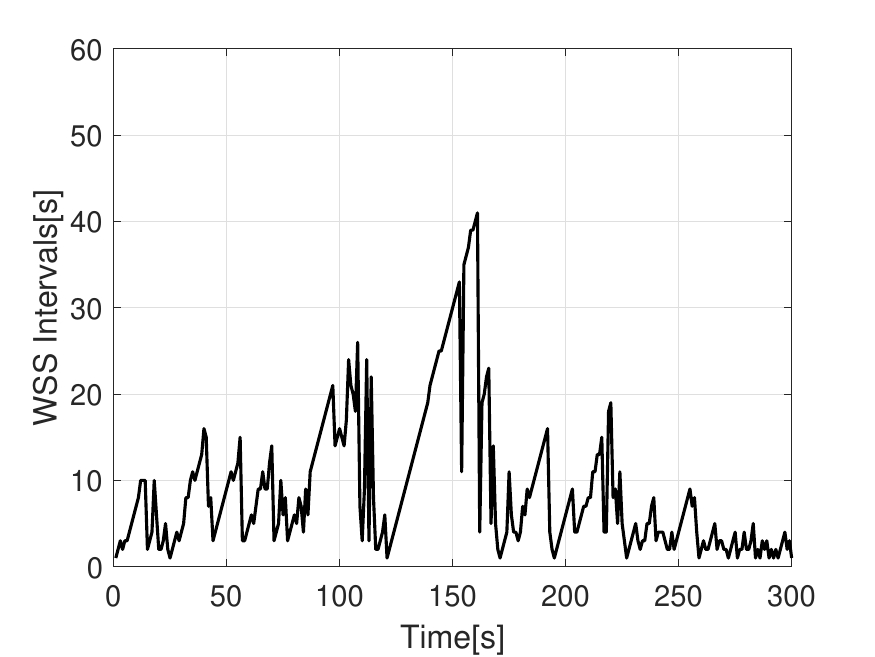}
    \caption{}
\end{subfigure}
\hspace{-5mm}
% \vspace{0.3cm}
\begin{subfigure}[t]{0.5\linewidth}
    \centering
    \includegraphics[width=\linewidth]{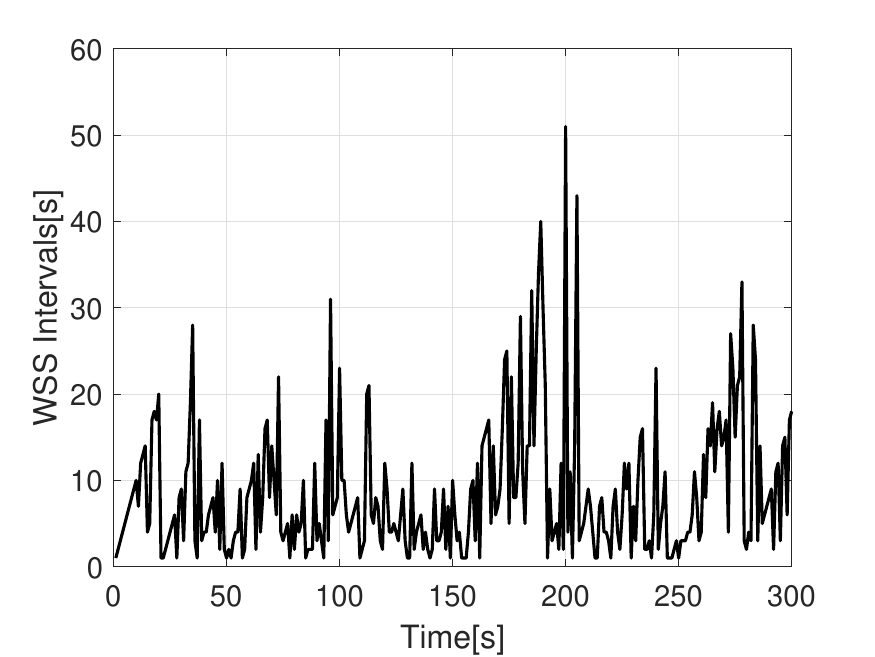}
    \caption{}
\end{subfigure}
\vspace{0.3cm}
\begin{subfigure}[t]{0.5\linewidth}
    \centering
    \includegraphics[width=\linewidth]{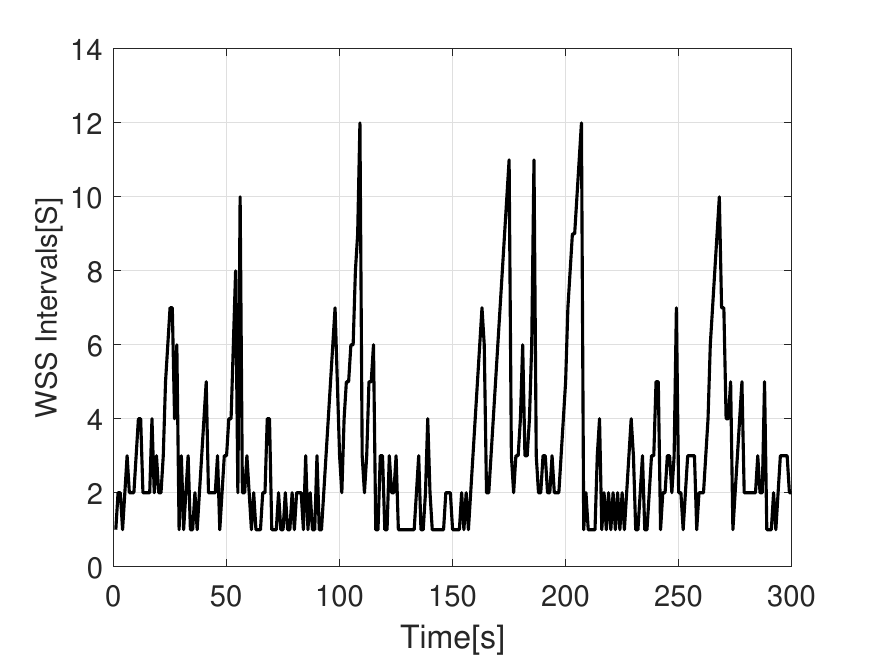}
    \caption{}
\end{subfigure}
\hspace{-5mm}
% \vspace{0.3cm}
\begin{subfigure}[t]{0.5\linewidth}
    \centering
    \includegraphics[width=\linewidth]{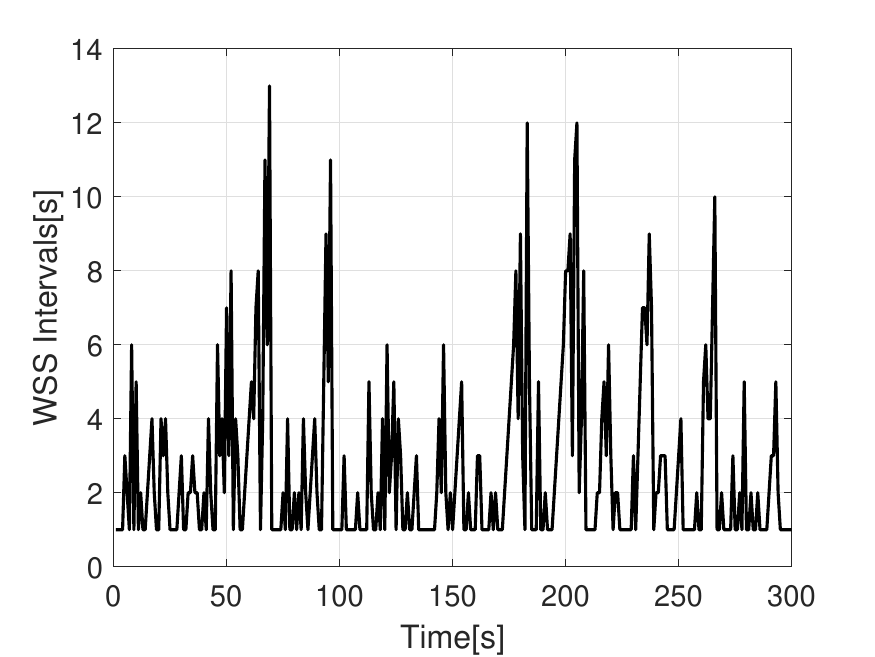}
    \caption{}
\end{subfigure}
\caption{Examples of RT simulated WSS intervals and the generated WSS intervals with the proposed CGAN-LSTM-SC in sparse and dense scenarios: (a) RT simulated WSS intervals in sparse scenarios, (b) CGAN-LSTM-SC generated WSS intervals in sparse scenarios, (c) RT simulated WSS intervals in dense scenarios, and (d) CGAN-LSTM-SC generated WSS intervals in dense scenarios.}
\label{WSS all}
\end{figure}

\begin{figure}[!t]
\centering
\begin{subfigure}[t]{0.9\linewidth}
    \centering
    \includegraphics[width=\linewidth]{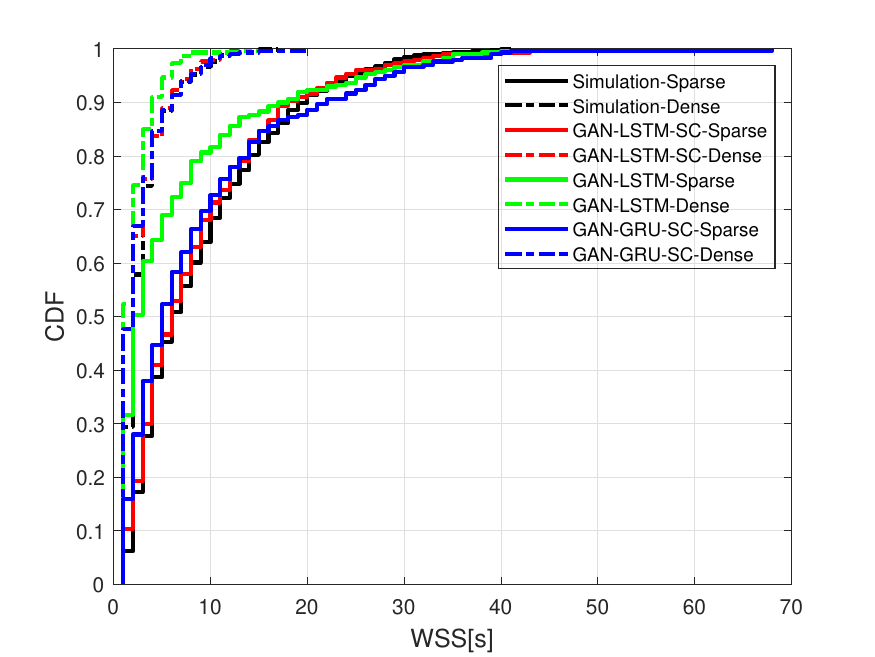}
    \caption{}
    \label{WSS}
\end{subfigure}
\hspace{-5mm}
\begin{subfigure}[t]{0.9\linewidth}
    %\centering
    \includegraphics[width=\linewidth]{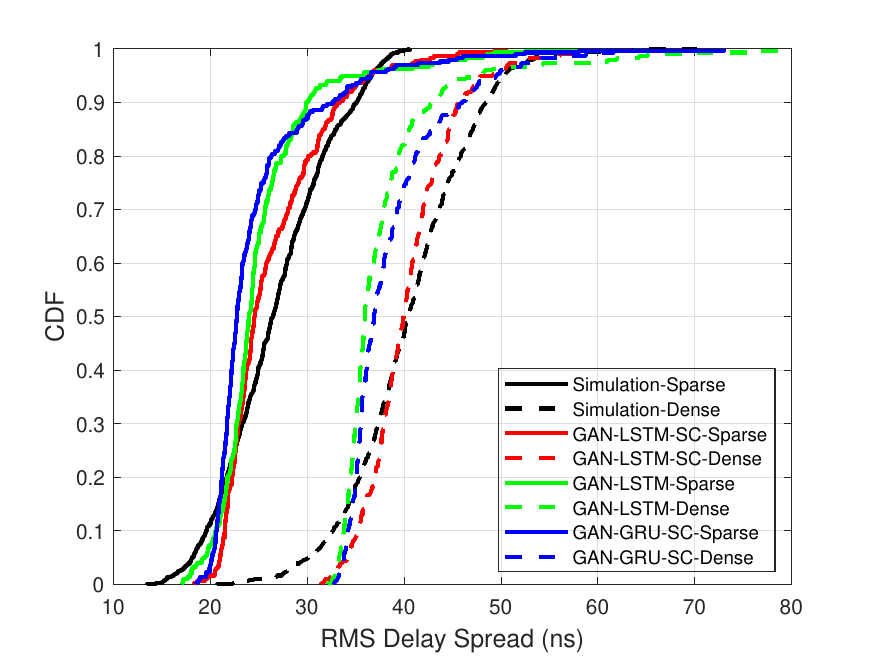}
    \caption{}
    \label{fig:RMSDS}
\end{subfigure}
\hspace{-5mm}
\begin{subfigure}[t]{0.9\linewidth}
    \centering
    \includegraphics[width=\linewidth]{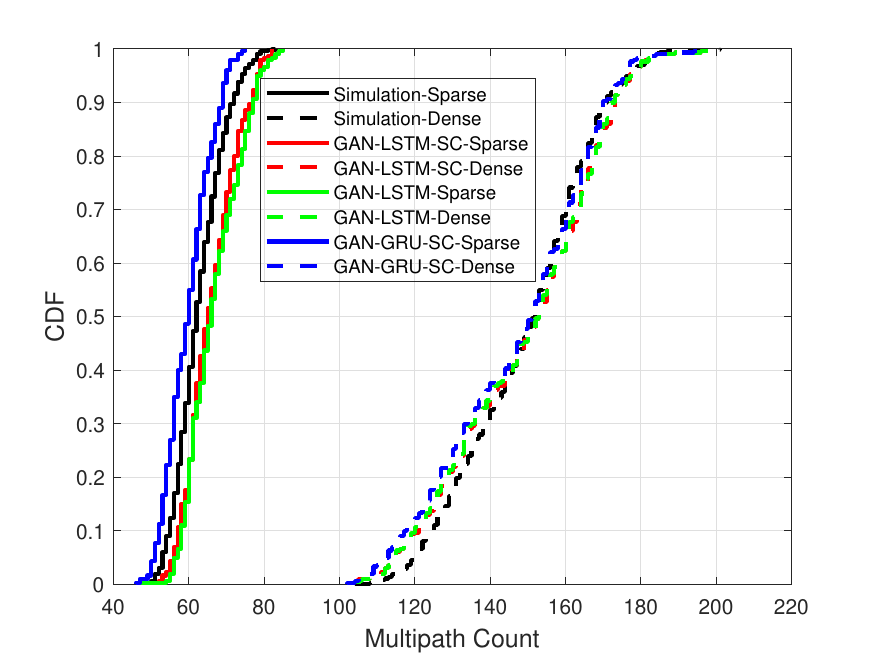}
    \caption{}
    \label{fig:MC}
\end{subfigure}
\caption{Comparison of distributions of non-stationary statistical features in sparse and dense scenarios: (a) WSS intervals, (b) RMSDS, and (c) multipath count.}
\label{ALL1}
\end{figure}

In this section, characteristics of the generated dynamic channels over time are analyzed by comparing PDPs with the simulated channels, and the model is validated by using statistical distribution of channel features. Existing methods for learning time-dependent characteristics of channel are also compared and analyzed.

\subsection{Channel Generation Performance}
To evaluate performance of channel generation, we compare the generated dynamic PDPs with RT simulations. The simulation PDPs are considered as ground truth and used for model training. The generated PDPs aim to replicate the dynamics in RT simulations while adhering to statistical distributions learned from the training data. By aligning the snapshots of dynamic PDPs from both sources, we observe in Fig. \ref{PDP} that the generated PDPs closely follows the temporal variations of the simulation PDPs.

\begin{figure}[!t]
\centering
\includegraphics[width=3.5in]{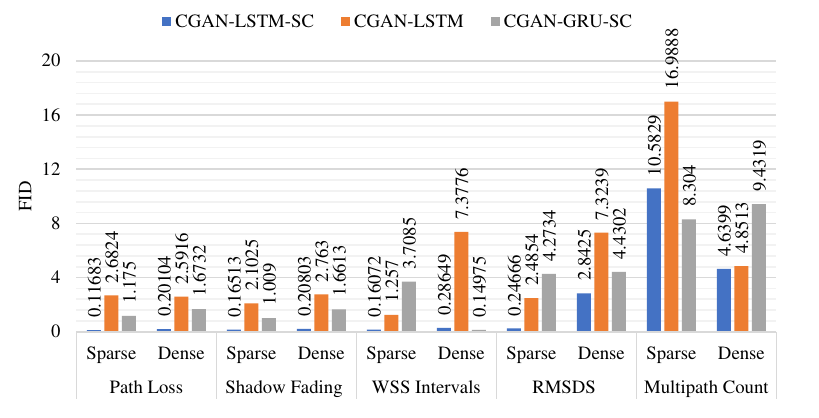}
\caption{Comparison of generation performance of different methods with different statistical parameters.}
\label{FID}
\end{figure}

\subsection{Channel Statistics Properties Comparison}
To comprehensively validate the accuracy of the generated channel, comparisons are conducted on the statistical properties of non-stationary channels. We compare the performance of four cases: RT simulations, GAN-LSTM-SC which is the proposed method with stationarity constraints, GAN-LSTM without stationarity constraints, and GAN-GRU-SC which replaces LSTM with GRU and has stationarity constraints. All the cases are compared under two types of scenarios: sparse and dense, which characterize weakly non-stationary and strongly non-stationary, respectively.

To ensure a thorough analysis, basic channel parameters such as path loss and shadow fading are compared in both sparse and dense scenarios. These comparisons serve as a baseline to verify consistency of the generated channels with the ground truth. Subsequently, non-stationary parameters are considered, including WSS intervals, RMSDS, and multipath count. These parameters are critical for capturing complex temporal and spatial dynamics of non-stationary channels.  Table \ref{channel statistics parameter} summarizes the comparisons of channel statistics in both sparse and dense scenarios. 

\subsubsection{Path Loss and Shadow Fading}

Path loss and shadow fading are typical large-scale channel parameters, and they both show similar characteristics for sparse and dense scenarios because of LOS propagation in the simulations. As shown in Fig. \ref{PL} and Fig. \ref{SF} , the path loss and shadow fading distributions generated by the proposed CGAN-LSTM-SC closely aligns with the simulation channels, demonstrating the model's ability to generate realistic distance-based propagation channels. The CGAN-LSTM and CGAN-GRU-SC also provide reasonable agreements with slightly higher errors.

\subsubsection{WSS Interval}
WSS interval is the key parameter used to characterize temporal non-stationary channel. To compare WSS distribution, we first calculate TPCC similarity between snapshots. Since the V2V scenario in this paper involves a fixed transmitter-receiver distance, LOS path remains constant. Therefore, we exclude the LOS path and calculate snapshot similarity for the remaining multipaths. Additionally,  correlation threshold is set to 0.7, considering two snapshots as belonging to the same WSS region only if their similarity is above this threshold. If similarity is below 0.7, the two snapshots are considered to be from different WSS regions. Fig. \ref{WSS all} illustrates the temporal variation of WSS intervals for two distinct stationary channels. The results indicate that WSS intervals are longer in sparse scenarios compared to dense scenarios. This difference shows that channel non-stationarity is more pronounced in dense scenarios. The increased non-stationarity can be attributed to the presence of a significantly richer set of scatterers in dense scenarios and the higher mobility of transceiver in these scenarios. 
In Fig. \ref{WSS}, it is found that the proposed framework performs the best, as it effectively captures temporal correlations and enforces constraints that ensure smoother transitions between snapshots. CGAN-GRU-SC performs reasonably well but it is less effective in capturing the fine-grained temporal dynamics of the channel, leading to slightly larger deviations. In contrast, CGAN-LSTM without stationarity constraints struggles to model the WSS characteristics properly due to the lack of temporal correlation modeling, resulting in the worst performance.

\subsubsection{RMSDS}
RMSDS quantifies degree of delay dispersion in channel. In sparse scenarios, RMSDS is relatively small. In contrast, in dense scenarios, RMSDS is large because of fluctuations of rich multipath propagation caused by scatterers. As shown in Fig \ref{fig:RMSDS}, there are differences in RMSDS of dynamic channels generated by different methods. The proposed CGAN-LSTM-SC closely matches simulation channels. Its ability to model temporal correlation in multipath variations ensures accurate replication of the RMSDS distribution. CGAN-GRU-SC generates reasonable RMSDS but struggles to fully capture the temporal complexity of multipath dynamics due to the simpler architecture of GRU. The RMSDS generated by baseline CGAN-LSTM deviates relatively significantly from simulation channels.

\subsubsection{Multipath Count}
Multipath count refers to number of distinct propagation paths received by receiver. In this study, signals with received power below -150 dB are considered noise. In sparse scatterer scenario, number of multipaths is relatively low due to fewer scatterers. In contrast, in dense scatterer scenario, number of multipaths increases due to a larger number of scatterers. Fig. \ref{fig:MC} shows a comparison of multipath count distributions in two different scenarios. All the three methods, i.e., CGAN-LSTM-SC, CGAN-LSTM, and CGAN-GRU-SC, show a robust ability to replicate the multipath count distributions. Each method effectively captures power distribution over different paths, ensuring consistency with the statistical characteristics of simulation channels.

 \subsection{Quantification of Similarity}
 In order to assess similarity between the channels generated under different methods and the simulation channels, similarity between the two distributions is quantified. The Fréchet Inception Distance (FID) is a metric used to evaluate quality of generated data by comparing its distribution to that of raw data\cite{heusel2017gans}, which is used to evaluate performance of the generated channel. The FID of raw and generated data is defined as follows:
 \begin{equation}
FID(x,g)=\|\mu_x-\mu_g\|^2+\mathrm{Tr}(\Sigma_x+\Sigma_g-2\sqrt{\Sigma_x\Sigma_g}),
\end{equation}
where $\mu_x$ and $\Sigma_x$ are the mean and covariance matrices of raw data. Parameters $\mu_g$ and $\Sigma_g$ are the mean and covariance matrix of the generated data. The lower the FID score, the closer the generated data is to the raw data in terms of both mean and covariance. In Fig. \ref{FID}, CGAN-LSTM-SC, CGAN-LSTM and CGAN-GRU-SC are compared. By calculating FID for each parameter, we evaluate effectiveness of each method in capturing statistical features of raw channel. Among them, the  proposed CGAN-LSTM-SC has the smallest FID for most of the parameters, indicating the best fit. The results demonstrate effectiveness of the proposed model in generating PDPs that closely resembles raw PDPs.

\section{Conclusion}

This paper presents a hybrid deep learning framework that integrates CGAN with LSTM networks to generate dynamic non-stationary channels. By incorporating stationarity constraints within generator and introducing supervision modules, the proposed method achieves high-precision dynamic channel modeling, effectively capturing complex temporal evolution and non-stationary properties of wireless channels. Additionally, this approach extends the applicability of existing channel measurement datasets and provides high-quality data generation with realistic non-stationary statistical features. The experimental results demonstrate that the statistical distribution and temporal dynamics between the generated and the raw channels are highly close in different levels of non-stationary scenarios.

\ifCLASSOPTIONcaptionsoff
  \newpage
\fi

% \bibliographystyle{ieeetr}
% \bibliography{paper}

\end{document}